\documentclass{ws-p10x7}
\usepackage{epsfig,eufrak}

\newcommand{\dds}{\stackrel{\leftrightarrow}{D}}

\begin{document}

\title{Structure Functions on the lattice}

\author{K. Jansen}

\address{CERN, TH Division, 1211 Geneva 23, Switzerland\\E-mail: Karl.Jansen@cern.ch}

\twocolumn[\maketitle\abstract{
We report on a lattice computation of the second moment of the pion matrix
element of the twist-2 non-singlet 
operator corresponding to the average momentum of parton densities. 
We apply a fully non-perturbatively evaluated running renormalization 
constant as well as a careful extrapolation of our results to the continuum limit.
Thus the only limitation of our final result is the quenched 
approximation. 
}]

\section{Introduction}

A reliable computation of parton distribution functions from first
principles would be very important for future experiments
planned e.g. at the LHC. The results of such calculations would offer
a unique way to test whether QCD is indeed the correct theory for the
strong interactions. 

In principle, the lattice regularization offers such a calculational scheme,
where the only starting point is the QCD lagrangian, and indeed
is able to give results for moments of parton density
distributions {\em in the continuum} and 
{\em fully non-perturbatively}, 
as we will show
in this contribution. 
The work presented here is a summary of a series of publications 
\cite{ref:perturbative,ref:non-pert,ref:universal,ref:invariant,ref:matrixele}
that provide the essential ingredients to reach the 
above ambitious aim.

The moments of parton density distributions
are related to expectation values of certain local
operators, which are renormalized multiplicatively by applying 
appropriate 
renormalization factors $Z(1/\mu)$ that depend
on the energy scale $\mu$.
This leads to consider renormalized 
matrix elements $O^{\mathrm{ren}}_{\mathrm{SF}}(\mu)$ to be  
computed in a certain, at this stage not specified, renormalization scheme SF. 
If the energy scale $\mu$ is chosen large enough, it is to be expected 
that the scale evolution is very well described by perturbation theory,
giving rise to the following definition  of
a {\em renormalization group invariant matrix element:}
\begin{equation}
O^{\mathrm{ren}}_{\mathrm{INV}} = O^{\mathrm{ren}}_{\mathrm{SF}}(\mu)\cdot
f^{\mathrm{SF}}(\bar{g}^2(\mu))
\label{eq:inv}
\end{equation}
with $\bar{g}(\mu)$ the running coupling and 
\begin{eqnarray}
  f^{\mathrm{SF}}(\bar{g}^2) & = & (\bar{g}^2(\mu))^{-\gamma_0/2b_0}
  \\
 & & \cdot \exp\left\{
    -\int_0^{\bar{g}(\mu)} dg\left[\frac{\gamma(g)}{\beta(g)} - \frac{\gamma_0}{b_0g} 
\right]\right\} \nonumber 
\label{eq:f}
\end{eqnarray}
where $\beta(g)$ and $\gamma(g)$ are the $\beta$ and anomalous-dimension functions
computed to a given order of perturbation theory in the specified scheme, i.e.
here the SF scheme. Once we know the value of $O^{\mathrm{ren}}_{\mathrm{INV}}$ evaluated
non-perturbatively, the running matrix element in a preferred scheme can be computed,
for example in the $\overline{\mathrm MS}$ scheme:
\begin{equation}
O^{\mathrm{ren}}_{\mathrm{\overline{\mathrm MS}}}(\mu) =  
O^{\mathrm{ren}}_{\mathrm{INV}}/f^{\overline{\mathrm{MS}}}(\bar{g}^2(\mu)) 
\end{equation}
with now, of course, the $\beta$ and $\gamma$ functions computed in the 
$\overline{\mathrm MS}$ scheme.

A non-perturbatively obtained value of the
renormalization group invariant matrix element is hence of central
importance.
Its calculation 
has to be performed in several steps. The reason is that we have to cover
a broad range of energy scales -- from the deep perturbative to the 
non-perturbative region. 
With the scale-dependent 
renormalization factor $Z(1/\mu)$ we can write the renormalized matrix element
of eq.~(\ref{eq:inv}) as 
\begin{equation}
O^{\mathrm{ren}}_{\mathrm{INV}} = \frac{\langle\pi | \mathcal{O}_{\mathrm{NS}}|\pi\rangle}
         {Z^{\mathrm{SF}}( 1/\mu )}
        \cdot f^{ \mathrm{SF} }(\bar{g}^2( \mu ))\; ,
\label{eq:split}
\end{equation}
with $\langle\pi | \mathcal{O}_{\mathrm{NS}}|\pi\rangle$ the expectation value
of the (non-singlet) operator under consideration in given states, here the pion states. 

So far, all our discussions have been in the continuum. However, if we think
of the lattice regularisation and eventual numerical simulations to obtain
non-perturbative results, it would be convenient to compute the renormalized matrix 
element at only one convenient (i.e. small hadronic) scale $\mu_0$.
We therefore rewrite the r.h.s. of eq.~(\ref{eq:split}) as:
\begin{equation}
           \frac{\langle\pi | \mathcal{O}_{\mathrm{NS}}|\pi\rangle}
           {Z^{\mathrm{SF}}( 1/\mu_0 )}
           \cdot \underbrace{\frac{Z^{\mathrm{SF}}( 1/\mu_0 )}
           {Z^{\mathrm{SF}}( 1/\mu )}}_{\equiv
           \sigma( \mu/\mu_0, \bar{g}(\mu)) }
           \cdot f^{ \mathrm{SF} }(\bar{g}^2( \mu ))
\end{equation}
where we 
introduce the {\rm step scaling function} $  \sigma( \mu/\mu_0, \bar{g}(\mu))$, which describes the 
evolution of the renormalization factor from a scale $\mu_0$ to a scale $\mu$. 
The advantage of concentrating on the step scaling function instead of the 
renormalization factor itself, is that the step scaling function is well defined
in the continuum and hence suitable for eventual continuum extrapolations
of lattice results. 

We finally write the r.h.s. of eq.~(\ref{eq:split}) as  
\begin{equation}
            O^{\mathrm{ren}}_{\mathrm{SF}}( \mu_0)
           \underbrace{\sigma( \mu/\mu_0, \bar{g}(\mu))
           \cdot f^{ \mathrm{SF} }(\bar{g}^2( \mu ))}_{
           \equiv \EuFrak{S}_{\rm INV}^{\rm UV}(\mu_0)}
\label{pieces_of_oren}
\end{equation}
with $O^{\mathrm{ren}}_{\mathrm{SF}}( \mu_0)$ the renormalized matrix element,
which is to be computed only once at 
a scale $\mu_0$ and the (ultraviolet) invariant step scaling function 
$\EuFrak{S}_{\rm INV}^{\rm UV}(\mu_0)$, which still depends on the infrared scale $\mu_0$. 
The following sections are devoted to a description of how these two basic ingredients can be
reliably computed on the lattice using non-perturbative methods, i.e. numerical
simulations.

\section{The renormalization group invariant step scaling function}

Let us start this section by disclosing what is hidden behind the fictitious SF 
scheme mentioned in the introduction. SF stands for Schr\"odinger functional
and denotes a finite physical volume, $V=L^3\cdot T$, renormalization
scheme where the energy scale
$\mu$ is identified with the inverse spatial length of the box itself, e.g. $\mu=1/L$.
The peculiarity of the Schr\"odinger functional set-up is that 
fixed boundary conditions in time $x_0$ are imposed with classical
fields at the time boundaries at time $x_0=0$ and $x_0=T$. For a more detailed discussion
we refer the reader to \cite{sf_general}. 

To discuss the renormalization of operators related to moments of parton
distribution functions, we first have to provide a renormalization 
condition. 
Denoting by $|SF\rangle$ a classical SF state, i.e. a classical quark field
at a time boundary with external momentum ${\bf p}$, the renormalization
condition that we will use reads
\begin{equation}
\langle SF | {\cal O}^\mathrm{ren}\left(\mu=\frac{1}{L}\right) | SF \rangle =
\langle SF | {\cal O}^{\rm tree} | SF \rangle\; .
\label{ren_cond}
\end{equation}
The relation between the expectation value of the bare operator and the 
renormalized one is established through a scale-dependent renormalization constant:

\begin{equation}
O^{R}(\mu) = Z^{-1}(1/\mu)O^{\rm bare}(1/L)\; .
\label{ren_operator}
\end{equation}
In perturbation theory, on the 1-loop level,
we have 
$Z(1/\mu) = 1 - \bar{g}^2(\mu) \left[ \gamma^{(0)} {\rm ln}(\mu) + B_0\right]$ with $\gamma^{(0)}$ 
the anomalous dimension and $B_0$ the constant part.
Up to this point, the discussion is given solely in the continuum where the SF renormalization 
scheme is a perfectly acceptable one. 
Different schemes such as the $\overline{\mathrm{MS}}$ scheme 
can be related to the SF scheme as usual 
in perturbation theory. 

If we are interested, however, in a non-perturbative calculation, we have to detour for a short 
time (which means, however, a substantial computer time) to a finite lattice with non-zero lattice spacing
$a$ that allows for numerical simulations.  A lattice representation of the twist-2 non-singlet
operator, which is the only case we are considering here, is given by 
\begin{equation}
{\cal O}_{12}(x) = \frac{1}{4}\bar{\psi}(x) \gamma_{\{1} \dds_{2\}} \frac{\tau^3}{2}\psi(x)\; ,
\label{eq:operator}
\end{equation}
\noindent where $\dds_{\mu}$ is the covariant derivative and the bracket
around indices means symmetrization.
The operator is probed by boundary quark fields $\zeta$ and $\bar{\zeta}$, which reside at $x_0=0$ and a 
correlation function is constructed
\begin{equation}
f_{O_{12}}(\frac{x_0}{a}) =  \sum_{\mathbf{x,y,z}} 
               \langle\rm{e}^{i\bf{p}(\bf{y}-\bf{z})}{\cal O}_{12}(x) \bar{\zeta}(y)\gamma_2 \tau^3\zeta(z)\rangle
\label{corr_12}
\end{equation}
with $\mathbf{p}$ the spatial 3-momentum. 

To take into account the effects of the boundary fields, we also consider the 
boundary operators
defined at the time boundaries $x_0=0$ and
$x_0=T$:
\begin{eqnarray}
{\cal O}_{0} & = & \frac{a^6}{L^3}\sum_{\bf{y},\bf{z}}
\bar\zeta({\bf{y}})\gamma_5 \frac{\tau^3}{2}\zeta({\bf{z}}), \nonumber \\  
{\cal O}_{T} & = & \frac{a^6}{L^3}\sum_{\bf{y},\bf{z}}
\bar\zeta'({\bf{y}})\gamma_5 \frac{\tau^3}{2}\zeta'({\bf{z}})
\label{boundary_ops}
\end{eqnarray}
from which we construct the correlation function:
\begin{equation}
f_{1} = -\langle {\cal O}_{0} {\cal O}_{T} \rangle\; .
\label{f1}
\end{equation}
The boundary wave-function contribution can then be taken out by considering the
ratio $f_{O_{12}}(x_0/a)/\sqrt{f_1}$. 

There are several physical scales in our problem which all have to
be given in units of $L$, which is the only scale 
that is to be changed. Therefore, in order 
to arrive at a definition of the renormalization constant, we have to specify
the spatial momentum 
$\mathbf{p}$, the quark mass $m_q$ and
the time $x_0$
where we read off the expectation value of the operator between SF states from 
the correlation function. 
The final physical
result does, of course, not depend on our choice of these quantities, but 
we choose them solely for convenience. In particular we select
\begin{eqnarray}
\label{conditions}
m_q & = & 0 \;,\;\;x_0 = T/4\; , \\
\mathbf{p} & \equiv & (p_1,p_2,p_3)=(p_1=2\pi/L,0,0)\; . \nonumber
\end{eqnarray}
The choice of a zero quark mass results in using a massless renormalization scheme
and the choice of the smallest available momentum on the lattice minimizes 
lattice artefacts. 
With the above choice, it is indeed only the physical box length $L$ 
(assuming $T=L$) that we identify with the
inverse scale, which is varied in the problem. 

We are now in a position to give the precise definition of the renormalization constant 
\begin{equation}
Z(L)=\bar{Z}(L)/Z_1(L)\; ,
\label{def_z}
\end{equation}
with
\begin{eqnarray}
\bar{Z}(L) & = & f_{O_{12}}(T/4)/f_{O_{12}}^\mathrm{tree}(T/4)\; ,\nonumber \\
\label{def_zbar}
Z_1(L) & = & \sqrt{f_1(L)}/\sqrt{f_1^\mathrm{tree}(L)}\; ,
\label{def_zf1}
\end{eqnarray}
where we divide by the corresponding tree level expression as required by
the renormalization condition, eq.~(\ref{ren_cond}).
Instead of computing the Z-factors, 
we concentrate on the 
step scaling functions 
\begin{equation}
\sigma_{\bar{Z}} = \frac{\bar{Z}(2L)}{\bar{Z}(L)},\;
       \sigma_{f_1} = \frac{Z_1(2L)}{Z_1(L)},\;
       \sigma_{Z} = \frac{Z(2L)}{Z(L)}\; ,
\label{def_ssf}
\end{equation}
because, 
in contrast to the Z-factors, the step scaling functions have a well-defined 
continuum limit. The strategy is now to compute the
step scaling functions at various values of the lattice spacing 
while keeping fixed the conditions in eqs.~(\ref{conditions}) and the physical
scale $\mu=L^{-1}$ (determined by the running coupling $\bar{g}(\mu)$)  
and to extrapolate the results thus obtained to $a=0$. 

\begin{figure}
\vspace{0.0cm}
\hspace{-0.0cm}
\psfig{file=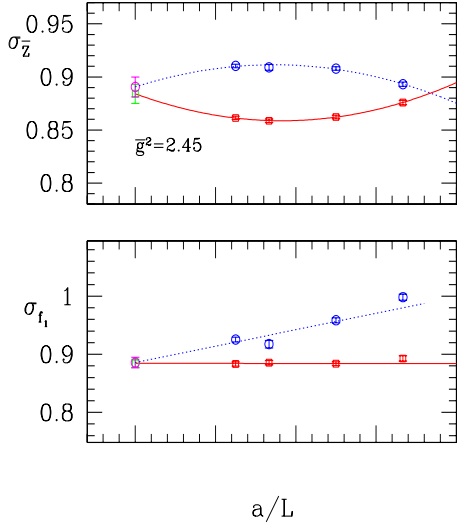,%
width=7cm,height=8cm}
\caption{ \label{fig:f1z}
Continuum extrapolation of the step scaling functions 
$\sigma_{\bar{Z}}$ and $\sigma_{f_1}$ performed 
separately for the Wilson action (circles and dotted lines) 
and the non-perturbatively improved action (squares and full lines)
at a fixed value of the running coupling $\bar{g}^2=2.45$.}
\end{figure}

\begin{figure}
\vspace{0.0cm}
\psfig{file=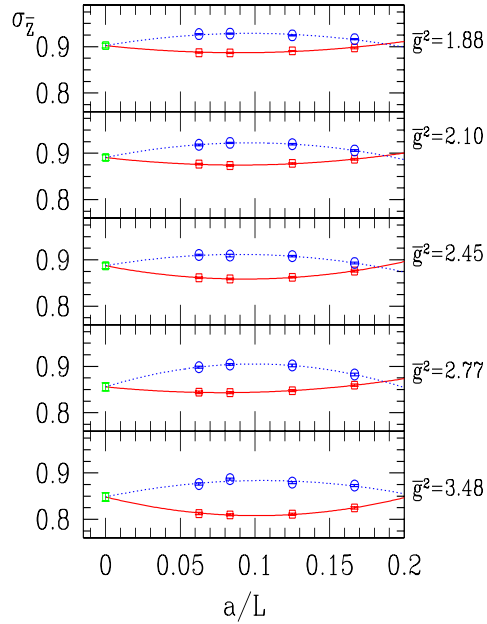,%
width=7cm,height=9.3cm}
\caption{ \label{fig:zbarcomb}
Constraint continuum extrapolation of $\sigma_{\bar{Z}}$ (notation as in fig.~\ref{fig:f1z}).}
\end{figure}

It is one of the basic ingredients and characteristics of our work
that almost all simulation results at non-zero lattice
spacings have been obtained by
employing
the standard
Wilson action and the non-perturbatively improved clover action.
Since these two formulations lead to different lattice artefacts,
it is a very crucial test of our results that their continuum extrapolations
give consistent results.
That this is indeed the case is demonstrated in fig.~\ref{fig:f1z}. It
shows that for the two step scaling functions $\sigma_{f_1}$ and
$\sigma_{\bar{Z}}$ the continuum limit of both discretizations agree within
the error bars. We note that in the case of $\sigma_{\bar{Z}}$ a quadratic
extrapolation in the lattice spacing $a$ is necessary while for
$\sigma_{f_1}$ a linear extrapolation is sufficient.
After checking that a similar behaviour is found at all values of
the coupling we have simulated, we performed constraint
fits,  demanding that the continuum value of the step scaling
functions be the same from both actions. A summary of our results for
$\sigma_{\bar{Z}}$ is shown in fig.~\ref{fig:zbarcomb}.

At this point, having performed all necessary continuum
extrapolations, we end our detour on a lattice with non-zero lattice
spacing and come back to the discussion in the continuum. With the 
results on the step scaling function, extrapolated to the continuum limit, which were obtained at 9 values
of the running coupling constant, we can now compute the (ultraviolet)
invariant step scaling function
\begin{equation}
  \EuFrak{S}_{\rm INV}^{\rm UV}(\mu_0) =
\sigma ( \mu/\mu_0 ,\bar{g}^2(\mu_0 ) )\cdot
                  f(\bar{g}^2(\mu ))\; ,
\label{rgi_ssf}
\end{equation}
which still depends on the infrared scale $\mu_0$. This scale dependence will
only be cancelled when multiplying with the matrix element,
renormalized at the scale $\mu_0$. The function $f(\bar{g}^2(\mu ))$ is the 
same as in eq.~(\ref{eq:f}) and the $\beta$ and $\gamma$ functions are taken 
up to 3 loops in the SF scheme\footnote{For the $\gamma$ function we have taken an
effective 3-loop parametrization as obtained by fitting our data to an 
effective 3-loop form \cite{ref:non-pert}.}.

\begin{figure}
\vspace{0.0cm}
\hspace{-0.0cm}
\psfig{file=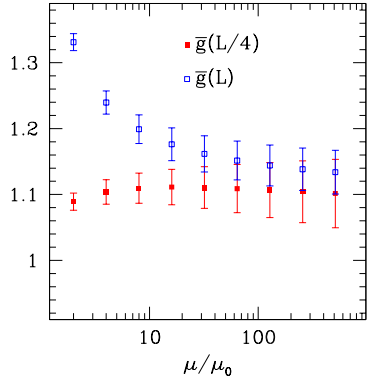,%
width=6cm,height=6cm}
\caption{ \label{fig:rgi_ssf}
The values of $\EuFrak{S}_{\rm INV}^{\rm UV}(\mu_0)$ for two choices
of the running coupling.}
\end{figure}

In fig.~\ref{fig:rgi_ssf} we show $\EuFrak{S}_{\rm INV}^{\rm UV}(\mu_0)$ as a function of
$\mu/\mu_0$. For large enough energy scales, $\mu/\mu_0 > 100$, 
$\EuFrak{S}_{\rm INV}^{\rm UV}(\mu_0)$ does not change within the errors and we can
determine a value for it by fitting the last, say, 4 points to a constant. 
Although 
there still is a scheme
dependence in the invariant step scaling function
through the remaining dependence on $\mu_0$, 
the value should be independent
of the choice of coupling used in the analysis. This is nicely illustrated
in fig.~\ref{fig:rgi_ssf}, where the choices of $\bar{g}(L/4)$ and $\bar{g}(L)$ 
give consistent values for the invariant step scaling function. 

We can therefore now give the first piece of information for the invariant matrix
element itself, as needed in eq.~(\ref{pieces_of_oren}), and quote
\begin{equation}
\EuFrak{S}_{\rm INV}^{\rm UV}(\mu_0)=1.11(4)\; .
\label{value_ssf}
\end{equation}

\section{Renormalized matrix element}

As a next step we have to compute the renormalized matrix element itself.
Again, we will always use the SF scheme and
we remark that this is the first time that a calculation of 
a 2-quark matrix element is attempted in
this set-up. 
We first tried to compute the matrix element (within pion states) of the
operator of eq.~(\ref{eq:operator}). However, since this operator needs a
non-vanishing momentum, we found, similar as in \cite{ref:schierholz}, that the signal is
very noisy. We therefore decided to switch to the operator

\begin{equation}
{\cal O}_{00}(x) = \bar\psi(x)\left[
\gamma_0 \dds_0 - \frac{1}{3}\sum_{k=1}^3 \gamma_k \dds_k\right]\frac{\tau^3}{8}\psi(x)
\label{eq:operator_44}
\end{equation}
which has the advantage that it can be computed at zero momentum. 
Taking the boundary operators of eq.~(\ref{boundary_ops})
we construct a correlation function

\begin{equation}
f_{\rm M}(x_0) = a^3\sum_{\bf{x}}
\langle {\cal O}_{0} {\cal O}_{00}(x) {\cal O}_{T} \rangle\; ,
\label{fm}
\end{equation}
which again is to be normalized by $f_1$, eq.~(\ref{f1}), to take out the boundary wave-function
contributions. 
Performing a transfer matrix decomposition, we find that for large enough values of $x_0$
and staying far enough from both boundaries
\begin{eqnarray}
f_1 &\simeq& \rho^2e^{-m_\pi T}, \nonumber \\
f_{\rm M}(x_0) &\simeq& \rho^2e^{-m_\pi T}\langle \pi | {\cal O}_{00} | \pi \rangle\; .
\end{eqnarray}
Assuming that there is a plateau region where
$f_M(x_0) / f_1 = \mathrm{const} \equiv \langle \pi | {\cal O}_{00} | \pi \rangle$,
and in which the first excited state
gives essentially no contributions, we obtain the physical matrix element $\langle x\rangle$ after
a suitable normalization (see \cite{ref:schierholz}):
\begin{equation}
\langle x \rangle \equiv \frac{2\kappa}{m_\pi}\langle \pi | {\cal O}_{00} | \pi \rangle\; .
\label{eq:ratio}
\end{equation}
In order to extract the matrix element of eq.~(\ref{eq:ratio}) we have chosen 
lattices with $T=3\;\mathrm{fm}$ and followed the
correlation function $f_{\rm M}(x_0)$ up to a distance of 1 fm in time direction.
At this distance we are sure that we project on the pion states as an inspection
of the pseudoscalar and axial-vector correlation functions (from which we also
extracted the pion masses) showed. Indeed, for $1\;\mathrm{fm} < x_0 < 2\;\mathrm{fm}$ 
the correlation function exhibits a plateau behaviour as can be seen in fig.~2 of 
ref.~\cite{ref:matrixele}.

Once we have the bare matrix element we need to renormalize it. To this end we computed
$Z(1/\mu_0)$ with $\mu_0^{-1}=1.436r_0$, $r_0\approx 0.5\;\mathrm{fm}$. We repeated such a calculation
for various lattice sizes, choosing the values of $\beta$ such that $L_{\mathrm{max}}=\mu_0^{-1}$
is kept fixed. Interpolating the numerical simulation data we obtain in this way
$Z(1/\mu_0)$ in a range of lattice spacing $0.05 \le a \le 0.1$, i.e. the range
of $a$ where the bare matrix element itself has been computed.
%

\begin{figure}
\vspace{0.0cm}
\hspace{-.0cm}
\psfig{file=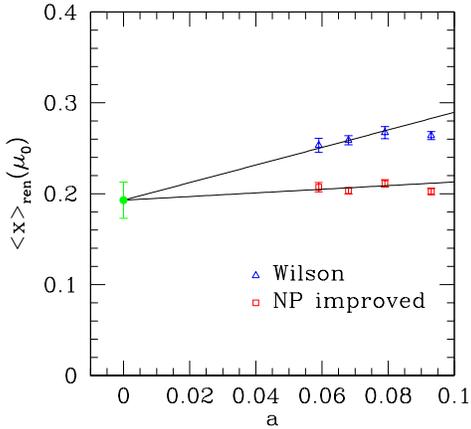,%
width=6.9cm,height=6.9cm}
\caption{ \label{fig:extra_12}
Constrained continuum extrapolation of the renormalized matrix element}
\end{figure}

This allowed us to apply the renormalization factor at exactly the same values
of $\beta$ where the matrix element has been computed. 
We show the matrix element renormalized with $Z_{12}(1/\mu_0)$ in 
fig.~\ref{fig:extra_12}. Again all calculations have been performed with two 
different discretizations, and we find quite convincingly the same
continuum limit. 

It might, however, not have escaped the reader's attention that we have used
the wrong renormalization factor, namely $Z_{12}$, for 
renormalizing 
the operator ${\cal O}_{00}(x)$ of eq.~(\ref{eq:operator_44}). 
The continuum extrapolation of the such renormalized operator needs a correction 
factor. 
In ref.~\cite{ref:matrixele} we have demonstrated that this correction
factor can be taken from perturbation theory and amounts to a shift of the
continuum renormalized matrix element by a few per cent. 
Taking the correction factor to be the same as found in \cite{ref:matrixele},
we can finally give our main result:
\begin{equation}
O^{\mathrm{ren}}_{\mathrm{INV}}=0.222(24)
\label{eq:oren_final}
\end{equation}

\section{Conclusion}

In this contribution we have demonstrated how we can compute on the lattice,
in a fully non-perturbative fashion, a renormalization group invariant
matrix element $O^{\mathrm{ren}}_{\mathrm{INV}}$ for the second moment of parton distributions of the
twist-2 non-singlet operator in the pion.
A preliminary value for $O^{\mathrm{ren}}_{\mathrm{INV}}$ is given in 
eq.~(\ref{eq:oren_final}) and can be used now as integration constant 
to obtain the renormalized matrix element at any scale in the preferred 
renormalization scheme. 

We want to emphasize that the value of the renormalized matrix element 
is shifted substantially from a value of about $\langle x\rangle(a=0.093)=0.30$ at $\beta=6$
to $\langle x\rangle(a=0)=0.2$ in the continuum limit. 
Hence we experience strong
lattice artefacts in the renormalized matrix element. 
Still, when the
matrix element is run in the $\overline{\mathrm{MS}}$ scheme to a scale
of $\mu=2.4\;\mathrm{GeV}$, we find $\langle x\rangle(\mu=2.4\mathrm{GeV})\approx 0.3$. 
Thus we find that, by a conspiration of two effects, our result agrees with
the number quoted in the pioneering work of \cite{ref:schierholz}. 
Therefore, 
the fact that within the quenched approximation, used here exclusively, the
number from the lattice simulations is higher than the experimental value
persists.

\vspace{-0.2cm}
\section*{Acknowledgements}
The work presented in this talk is part of a most enjoyable
and lively 
collaboration with 
M.~Guagnelli and R.~Petronzio. I also want to thank 
A.~Bucarelli, F.~Palombi, and A.~Shindler for 
useful discussions
and essential contributions.

\def\NPB #1 #2 #3 {Nucl.~Phys.~{\bf#1} (#2)\ #3}
\def\NPBproc #1 #2 #3 {Nucl.~Phys.~B (Proc. Suppl.) {\bf#1} (#2)\ #3}
\def\PRD #1 #2 #3 {Phys.~Rev.~{\bf#1} (#2)\ #3}
\def\PLB #1 #2 #3 {Phys.~Lett.~{\bf#1} (#2)\ #3}
\def\PRL #1 #2 #3 {Phys.~Rev.~Lett.~{\bf#1} (#2)\ #3}
\def\PR  #1 #2 #3 {Phys.~Rep.~{\bf#1} (#2)\ #3}

\def\etal{{\it et al.}}
\def\ibid{{\it ibid}.}

\end{document}